\documentclass[a4paper]{article}

\usepackage{INTERSPEECH2019}

\newcommand{\bs}[1]{\boldsymbol{#1}}

\usepackage{hyperref}
\usepackage{subfigure}

\title{Training a Neural Speech Waveform Model using Spectral Losses of Short-Time Fourier Transform and Continuous Wavelet Transform}
\name{Shinji Takaki$^1$, Hirokazu Kameoka$^2$, Junichi Yamagishi$^{1,3}$}
\address{$^1$National Institute of Informatics, Japan. \\
$^2$NTT Communication Science Laboratories, Nippon Telegraph and Telephone Corporation, Japan.\\
$^3$The Centre for Speech Technology Research, University of Edinburgh, United Kingdom.
}
\email{takaki@nii.ac.jp, hirokazu.kameoka@lab.ntt.co.jp, jyamagis@nii.ac.jp}

\begin{document}

\maketitle
\begin{abstract} 
Recently, we proposed short-time Fourier transform (STFT)-based loss functions for training a neural speech waveform model. In this paper, we generalize the above framework and propose a training scheme for such models based on spectral amplitude and phase losses obtained by either STFT or continuous wavelet transform (CWT), or both of them. Since CWT is capable of having time and frequency resolutions different from those of STFT and is cable of considering those closer to human auditory scales, the proposed loss functions could provide complementary information on speech signals. Experimental results showed that it is possible to train a high-quality model by using the proposed CWT spectral loss and is as good as one using STFT-based loss.
\end{abstract}

\section{Introduction}
The performance improvements of speech synthesis systems brought by recent neural speech waveform models have been remarkable. Various high-quality models such as WaveNet \cite{oord2016wavenet}, SampleRNN \cite{mehri2016samplernn}, Parallel WaveNet \cite{oord2017parallel}, WaveGlow \cite{prenger2018waveglow}, LPCNet \cite{valin2018lpcnet}, and LP-WaveNet \cite{hwang2018lp} have been proposed. These models are frequently used as vocoders to convert acoustic features, e.g., the mel spectrogram, into speech waveforms. Such neural vocoders are an essential component in various speech synthesis applications.

One open question of such models is their training procedure. Several training methods and criteria for these models have been studied. To give a few examples, it was initially proposed to train a neural waveform model as a classification model by using a quantized speech waveform \cite{oord2016wavenet}. This method was improved on the basis of a discretized mixture of logistic distributions \cite{oord2017parallel}. Distillation from an auto-regressive (AR) network to a non-AR network was also investigated \cite{oord2017parallel,ping2018clarinet}. Normalizing Flows and Glow were also investigated to consider invertible transformations from an AR structure to a non-AR structure \cite{DBLP:journals/corr/abs-1811-02155,prenger2018waveglow}. A combination of linear AR (linear predictive coding) and non-linear AR was also investigated \cite{Juvela2018,valin2018lpcnet,hwang2018lp}. 

\begin{figure}[t]
  \centering
  \includegraphics[width=1.0\columnwidth]{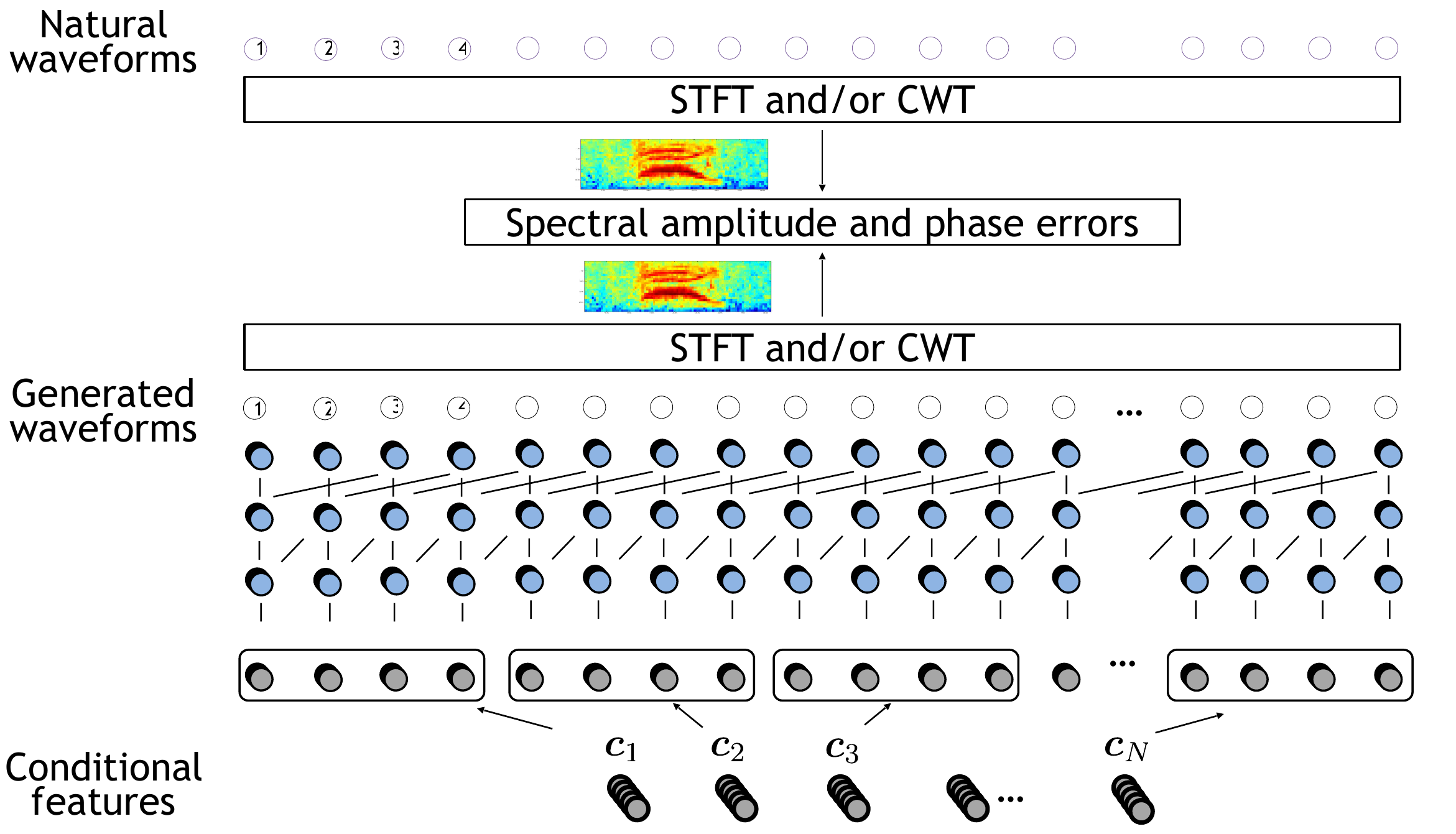}
  \caption{Overview of proposed CWT-based loss for neural speech waveform models}
  \label{fig:CWT-waveform}
\end{figure}

Recently, we proposed a new loss defined in the short-time Fourier transform (STFT) time-frequency domain for training neural speech waveform models \cite{takaki2018stft} as shown in Figure \ref{fig:CWT-waveform}. The loss can be used for efficiently training a model \textit{without using a time-consuming AR structure} because the STFT spectrum can contain multiple speech waveform samples and because a waveform model can be explicitly optimized on the basis of spectral amplitude and phase information \cite{wang2018neural}.

In this paper, we generalize the above framework and propose a new training scheme for neural speech waveform models based on spectral amplitude and phase losses obtained by either STFT or continuous wavelet transform (CWT), or both of them. In STFT, time and frequency resolutions are determined by the shape of the analysis window and frame shift, and it is not straightforward to balance them. Time-frequency analysis with different temporal and frequency resolutions can be achieved by using the CWT. It is also possible to adjust the CWT's time-frequency analysis so that we can consider scales similar to human auditory scales. In this paper, we investigate the training of a high-quality neural speech waveform model by using both STFT and CWT spectral losses.

The rest of this paper is structured as follows. In Section 2, we overview CWT. We explain the proposed loss in Section 3 and report experimental conditions and results in Section 4 as a proof of concept. We summarize our findings in Section 5.   

\section{Continuous wavelet transform (CWT)}
\label{sec:cwt}

\begin{figure}[t]
  \centering
  \includegraphics[width=1.0\columnwidth]{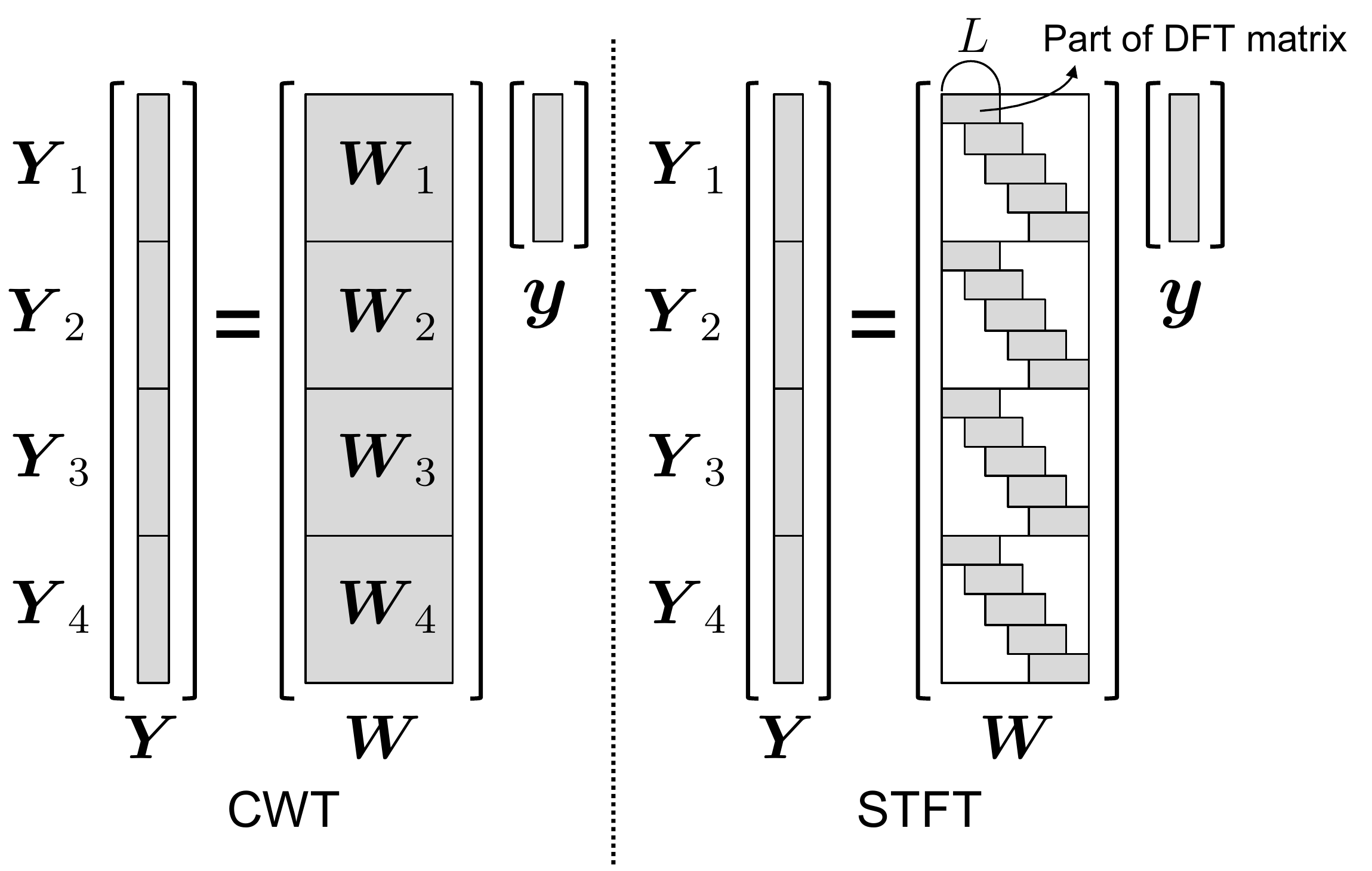}
  \vspace{-2mm}
  \caption{CWT complex spectra calculated by using matrix $\bs{W}$. Here, $\bs{W}\in\mathbb{C}^{LT\times M}$ and indexes represent CWT operation and scales. In STFT case, scale indexes and $L$ denote frequency bins and frame length. White parts in matrix $\bs{W}$ represent $0$.}
  \label{fig:CWT}
  \vspace{-5mm}
\end{figure}

In this paper, CWT for discrete signals $\bs{y}$ is represented by using CWT matrix $\bs{W}\in\mathbb{C}^{LT \times T}$ as shown in Figure \ref{fig:CWT} \cite{tomohikoDAFx2014}:
\begin{align}
\label{eq:cwt}
\bs{Y} &= \bs{W}\bs{y}, \\
\bs{W} &= \left[ \begin{array}{c} \bs{W}_0 \\ \bs{W}_1 \\ \vdots \\ \bs{W}_{L-1} \end{array} \right],
\bs{W}_l = \left[ \begin{array}{cccc} \psi_{l,0} & \psi_{l,1} & \dots & \psi_{l,T-1} \\ \psi_{l,T-1} & \psi_{l,0} & \dots & \psi_{l,T-2} \\ \vdots & \vdots & \ddots & \vdots \\ \psi_{l,1} & \psi_{l,2} & \dots & \psi_{l,0} \\ \end{array} \right],
\end{align}
where $l$ and $t$ represent a scale and time shift parameter in the continuous wavelet transform. Also, $\psi_{l,t} = \psi(t\delta/a_{l})$ is a scaled mother wavelet, and $a_{l}$ and $\delta$ represent a scale parameter and a signal sampling interval, respectively. Since a special case of the CWT matrix $\bs{W}$ includes a transformation to perform STFT \cite{tomohikoDAFx2014}, we use Eq. (\ref{eq:cwt}) to represent both STFT and CWT in the following sections.

\begin{figure}[t]
  \begin{center}
  \subfigure[STFT spectrogram]{\includegraphics[width=1.0\columnwidth]{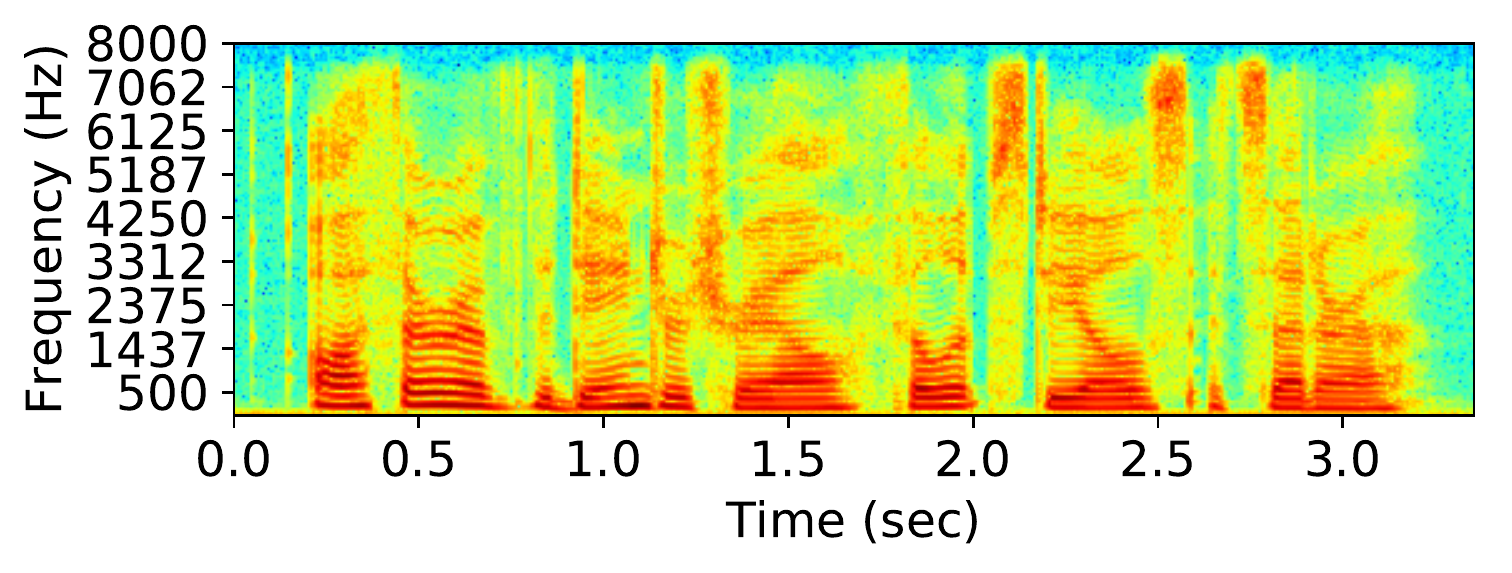}}
  \subfigure[CWT spectrogram with mel-frequency scale]{\includegraphics[width=1.0\columnwidth]{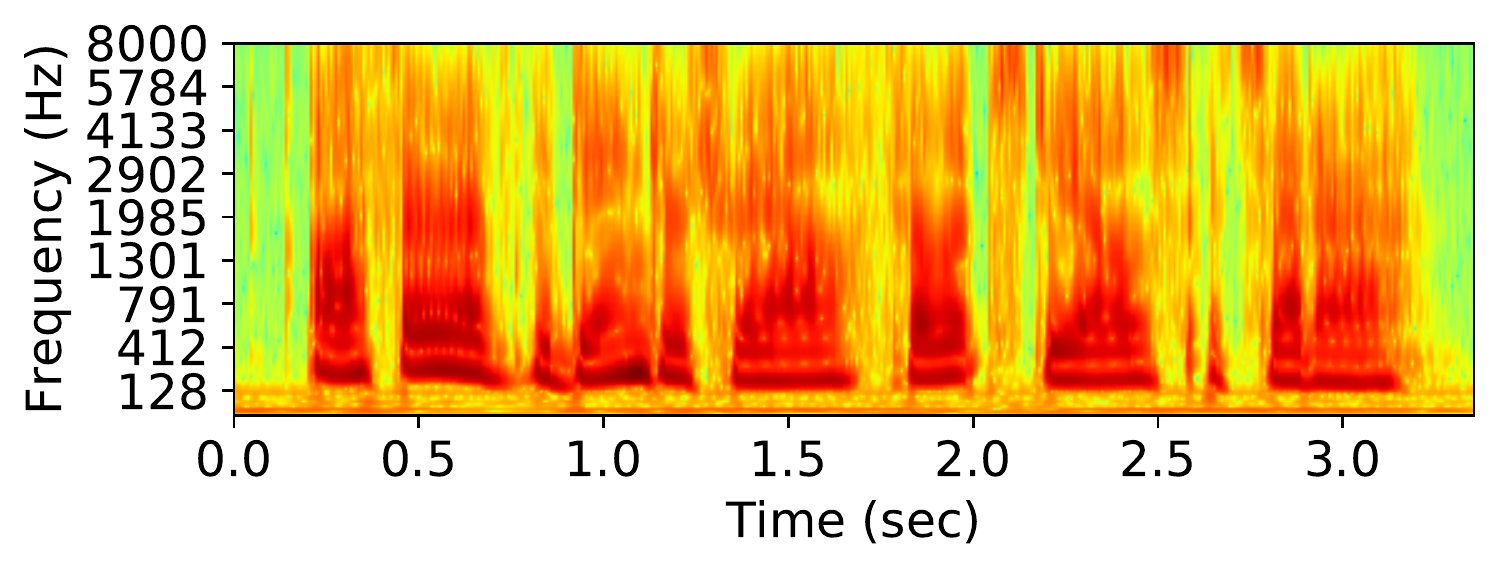}}
  \subfigure[CWT spectrogram with log-frequency scale]{\includegraphics[width=1.0\columnwidth]{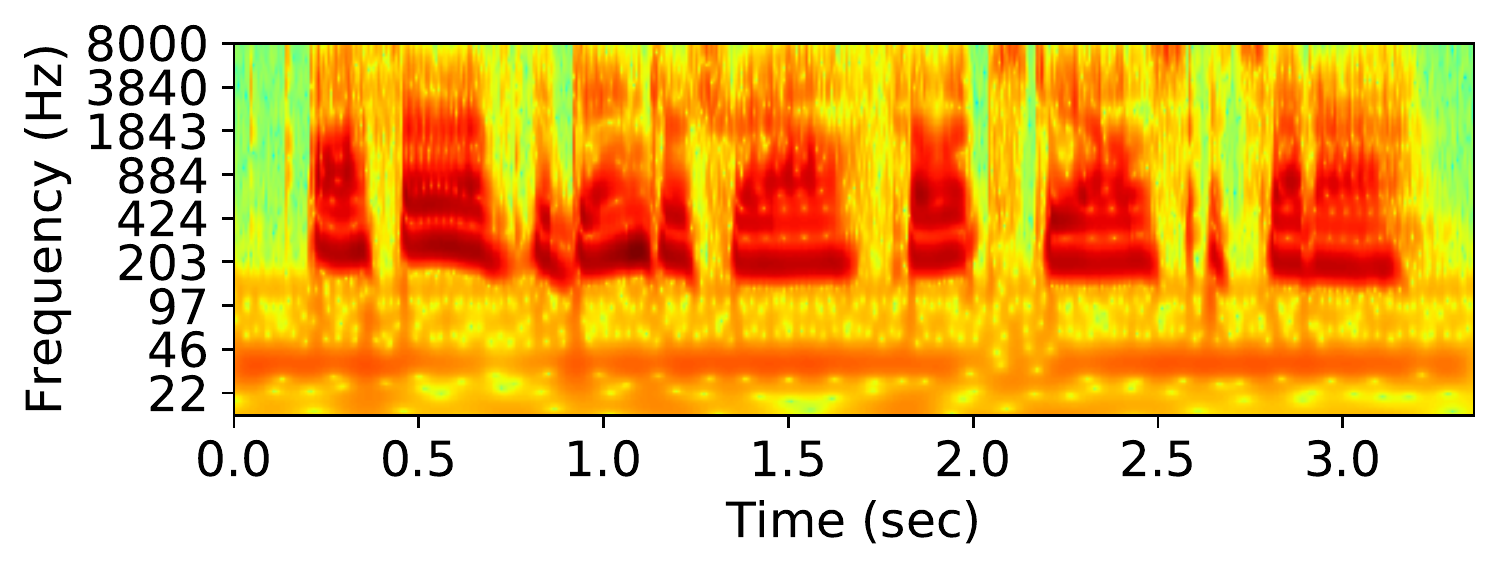}}
  \end{center}
  \caption{STFT spectrogram (a) and CWT spectrograms whose frequency resolution are equally spaced on mel-frequency scale (b) and log-frequency scale (c). 0th is not included, and total number of scales is 256.}
  \label{fig:stft_logcwt_melcwt}
\end{figure}

CWT can analyze a speech waveform with a time-frequency resolution different from that of STFT. Fig.\ \ref{fig:stft_logcwt_melcwt} shows an STFT spectrogram and CWT spectrograms in which, from top to bottom, the frequency resolutions are on a linear-frequency scale, mel-frequency scale, and log-frequency scale. 

%
%

\section{CWT spectral loss}

As described in Section \ref{sec:cwt}, both STFT and CWT spectra can be represented by Eq. (\ref{eq:cwt}). This provides a unified way of defining both STFT spectral loss and CWT spectral loss in order to train a neural speech waveform model. We define amplitude and phase spectral losses the same as those proposed in \cite{takaki2018stft}. An STFT/CWT complex spectrum $Y_{l,t}$, amplitude spectrum $A_{l,t}$, and phase spectrum $\exp(\theta_{l,t})$ (to which the Euler formula is applied) of scale $l$ at time $t$ are represented by using the $t$-th row vector of $\bs{W}_l$:
\begin{align}
\label{eq:complex}
Y_{l,t} &= \bs{W}_{l,t} \bs{y}, \\
\label{eq:amplitude}
A_{l,t} &= |Y_{l,t}| \nonumber \\
        &= (\bs{y}^{\top}\bs{W}_{l,t}^{H}\bs{W}_{l,t} \bs{y})^{\frac{1}{2}}, \\
\label{eq:phase}
\exp(i\theta_{l,t}) &= \exp(i\angle Y_{l,t}) \nonumber \\
&= \frac{Y_{l,t}}{A_{l,t}} \nonumber \\
&= \frac{\bs{W}_{l,t} \bs{y}}{(\bs{y}^{\top}\bs{W}_{l,t}^{H}\bs{W}_{l,t} \bs{y})^{\frac{1}{2}}}.
\end{align}

Amplitude and phase spectral losses used for training a neural speech waveform model are defined by Eqs.\ (\ref{eq:amplitude}) and (\ref{eq:phase}). Please see \cite{takaki2018stft} for details on the derivatives.

\subsection{Amplitude spectral loss}
The amplitude spectral loss is defined as follows.
\begin{align}
\label{eq:err_amp}
E^{(A)}_{l,t} &= \frac{1}{2} (\hat{A}_{l,t} - A_{l,t})^2 \\
&= \frac{1}{2} (\hat{A}_{l,t} - (\bs{y}^{\top}\bs{W}_{l,t}^{H}\bs{W}_{l,t} \bs{y})^{\frac{1}{2}})^2,
\end{align}
where $\cdot^H$ represents a Hermitian transpose. A partial derivative of $E^{(A)}_{l,t}$ with respect to $\bs{y}$ is as follows.
\begin{align}
\label{eq:pd_amp}
\frac{\partial{E^{(A)}_{l,t}}}{\partial{\bs{y}}} &= \left(A_{l,t} - \hat{A}_{l,t}\right) \mathcal{R}\left(\exp(i\theta_{l,t})\bs{W}^{H}_{l,t}\right),
\end{align}
where $\mathcal{R}(z)$ represents a real part of a complex number $z$.

\subsection{Phase spectral loss}
The phase spectral loss is defined as follows to consider this periodic property.
\begin{align}
\label{eq:err_ph}
E^{(P)}_{l,t} &= \frac{1}{2} \left|1 - \exp(i(\hat{\theta}_{l,t}-\theta_{l,t})) \right|^2 \\
&= 1 - \frac{1}{2} \left(\frac{\hat{Y}_{l,t}}{\hat{A}_{l,t}}\frac{(\bs{y}^{\top}\bs{W}_{l,t}^{H}\bs{W}_{l,t}\bs{y})^{\frac{1}{2}}}{\bs{W}_{l,t}\bs{y}}\right.\nonumber\\
&\hspace{10mm}\left . + \frac{\overline{\hat{Y}}_{l,t}}{\hat{A}_{l,t}}\frac{(\bs{y}^{\top}\bs{W}_{l,t}^{H}\bs{W}_{l,t} \bs{y})^{\frac{1}{2}}}{\overline{\bs{W}}_{l,t}\bs{y}}\right),
\end{align}
where $\overline{\cdot}$ represents a complex conjugate. A partial derivative of $E^{(P)}_{l,t}$ with respect to $\bs{y}$ is as follows.
\begin{align}
\label{eq:pd_ph}
\frac{\partial{E^{(P)}_{l,t}}}{\partial{\bs{y}}} &=
\sin(\hat{\theta}_{l,t} - \theta_{l,t}) \mathcal{I}\left(\frac{1}{\overline{Y}_{l,t}}\bs{W}_{l,t}^{H}\right),
\end{align}
where $\mathcal{I}(z)$ represents an imaginary part of a complex number $z$. 

\subsection{Proposed loss function for model training}
In this paper, the proposed loss function used for model training $E$ is represented by the weighted sum of an STFT amplitude loss $E_{l,t}^{(A, FT)}$, STFT phase loss $E_{l,t}^{(P,FT)}$, CWT amplitude loss $E_{l,t}^{(A,WT)}$, and CWT phase loss $E_{l,t}^{(P,WT)}$.
\begin{align}
\label{eq:loss}
E &= \sum_{l,t} \left(\alpha_{l,t}^{(A, FT)} E_{l,t}^{(A, FT)} + \alpha_{l,t}^{(P, FT)} E_{l,t}^{(P,FT)}\right) \nonumber \\ 
  &  \hspace{5mm} + \sum_{l',t'} \left(\alpha_{l',t'}^{(A, WT)} E_{l',t'}^{(A,WT)} + \alpha_{l',t'}^{(P, WT)} E_{l',t'}^{(P,WT)}\right),
\end{align}
where $\alpha_{l,t}^{(A, FT)}$, $\alpha_{l,t}^{(P, FT)}$, $\alpha_{l',t'}^{(A, WT)}$, and $\alpha_{l',t'}^{(P, WT)}$ are weights for STFT amplitude and phase spectral losses and CWT amplitude and phase spectral losses, respectively.

\section{Experiments}
Neural vocoders were trained by utilizing mel spectrograms as conditional features on the basis of the proposed loss. We conducted analysis-by-synthesis experiments\footnote{Synthetic speech samples can be found at \url{https://nii-yamagishilab.github.io/samples-STFTCWT/index.html}}.

\subsection{Experimental condition}
\begin{figure}[t]
  \centering
  \includegraphics[width=1.0\columnwidth]{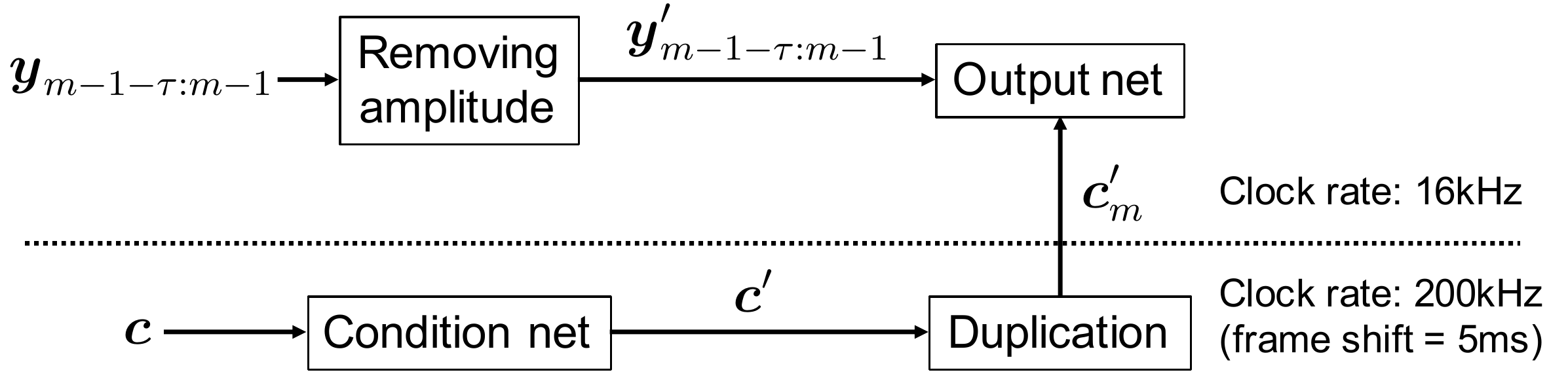}
  \caption{Network architecture of neural vocoders used in experiment}
  \label{fig:network}
\end{figure}
\begin{table}[t]
  \begin{center}
    \begin{tabular}{|c|c|c|}
      \hline
      Name & Conditions & Weights \\ \hline \hline
      WORLD & mgc, F0, bap & - \\ \hline
      STFT & & $\alpha_{l,t}^{(A, FT)} = 1$, $\alpha_{l',t'}^{(A, WT)} = 0$ \\ \cline{1-1}\cline{3-3}
      STFT+CWT & Mel spec.\  & $\alpha_{l,t}^{(A, FT)} = 0.5$, $\alpha_{l',t'}^{(A, WT)} = 0.5$ \\ \cline{1-1}\cline{3-3}
      CWT & & $\alpha_{l,t}^{(A, FT)} = 0$, $\alpha_{l',t'}^{(A, WT)} = 1$ \\ \hline
      \end{tabular}
    \end{center}
    \caption{List of evaluated vocoders. Here, WORLD, mgc, and bap mean WORLD vocoder \cite{morise2016world}, mel cepstrum coefficients, and band aperiodic components, respectively. FT, FT+WT, and WT were neural vocoders and were trained on basis of different weights.}
    \label{tab:method}
\end{table}
\textbf{Database} We used $1,032$ utterances as training data and $50$ utterances as test data of a female speaker, SLT, from the CMU-ARCTIC database. The sampling rate was $16$ kHz.

\vspace{2mm}
\noindent
\textbf{Network} The same architecture as that used in \cite{takaki2018stft}, shown in Fig. \ref{fig:network}, was used as a neural vocoder. Conditional features were first converted into hidden representations through a conditional network consisting of an $80$-unit bi-directional LSTM and a CNN with $80$ filters whose size was $5$ (time direction) $\times$ $80$ (frequency direction). Then, hidden representations were duplicated to adjust the time resolution. The previous $400$ samples and hidden representations obtained from the conditional network were fed into an output network. The output network was composed of three $256$-unit uni-directional LSTMs. As with \cite{takaki2018stft}, waveform samples obtained by removing amplitude information were used as feedback. Mini-batches were created, each from $120$ randomly selected speech segments. Each mini-batch contained speech waveform samples that each totaled $15$ s (each speech segment equaled $0.125$ s). We used the Adam optimizer \cite{DBLP:journals/corr/KingmaB14}, and the number of updating iterations was $80$k.

\vspace{2mm}
\noindent
\textbf{Conditional features}
An 80-dim.\ mel spectrogram was used as the conditional feature of the neural vocoder. To extract a mel spectrogram, the frame shift, frame length, and FFT size were set to $80$, $400$, and $512$, respectively. 

\vspace{2mm}
\noindent
\textbf{Spectral loss}
The frame shift, frame length, and FFT size were $1$, $400$, and $512$ respectively, in the calculation of the STFT spectral loss. Equal intervals on the mel-frequency scale, $\omega=6$, and a complex Morlet wavelet were used to calculate the CWT spectral loss. The total number of scales for CWT was set to $25$ when a combination of STFT and CWT spectral losses was used and to $257$ when only CWT spectral loss was used.

\vspace{2mm}
\noindent
\textbf{Weights of loss function}
From preliminary experimental results, it was found that a noisy speech waveform was generated when CWT phase spectral loss was used. Thus, we omitted this loss from the training criterion $E$, i.e., $\alpha_{l',t'}^{(P, WT)}=0$. Furthermore, voiced/unvoiced flags (1: voiced, 0: unvoiced) were used as weights for the STFT phase spectral loss, i.e., $\alpha_{t,n}^{(P, FT)} = v_{t}$, because \cite{takaki2018stft} has shown that voiced/unvoiced flags were useful in training a neural speech waveform model. The tested weights for STFT and CWT amplitude spectral losses were as follows.
\begin{itemize}
\item  \textbf{$\alpha_{l,t}^{(A, FT)} = 1$, $\alpha_{l',t'}^{(A, WT)} = 0$: }Only STFT amplitude spectral loss was used as in the conventional training.

\item \textbf{$\alpha_{l,t}^{(A, FT)} = 0.5$, $\alpha_{l',t'}^{(A, WT)} = 0.5$: }A combination of STFT amplitude spectral loss and CWT amplitude spectral loss was used.

\item  \textbf{$\alpha_{l,t}^{(A, FT)} = 0$, $\alpha_{l',t'}^{(A, WT)} = 1$: }Only CWT amplitude spectral loss was used.
\end{itemize}

In the experiment, we compared the above neural vocoders with the WORLD vocoder. We extracted spectral envelopes, F0, and aperiodic components, and they were converted into low-dimensional features of $59$-dim.\ mel cepstrum coefficients and 21-dim.\ band aperiodic components. During the synthesis phase, the low-dimensional features were converted into original high-dimensional features and input to the WORLD vocoder.

\begin{figure*}[t]
  \centering
  \subfigure[NAT]{\includegraphics[height=5.2cm]{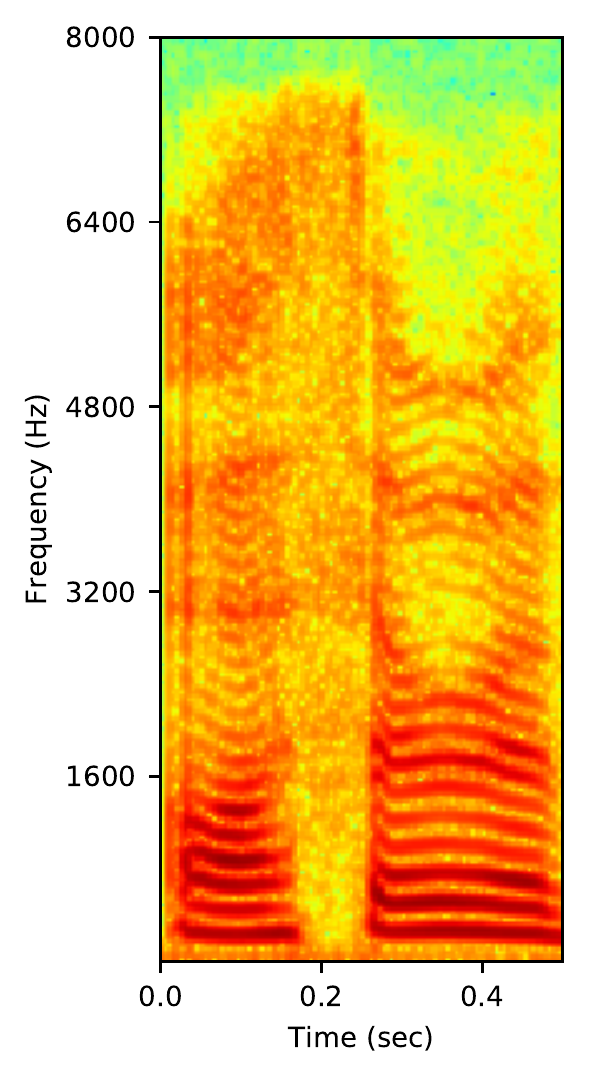}\label{fig:NAT_stft_spectrogram}}
  \subfigure[WORLD]{\includegraphics[height=5.2cm]{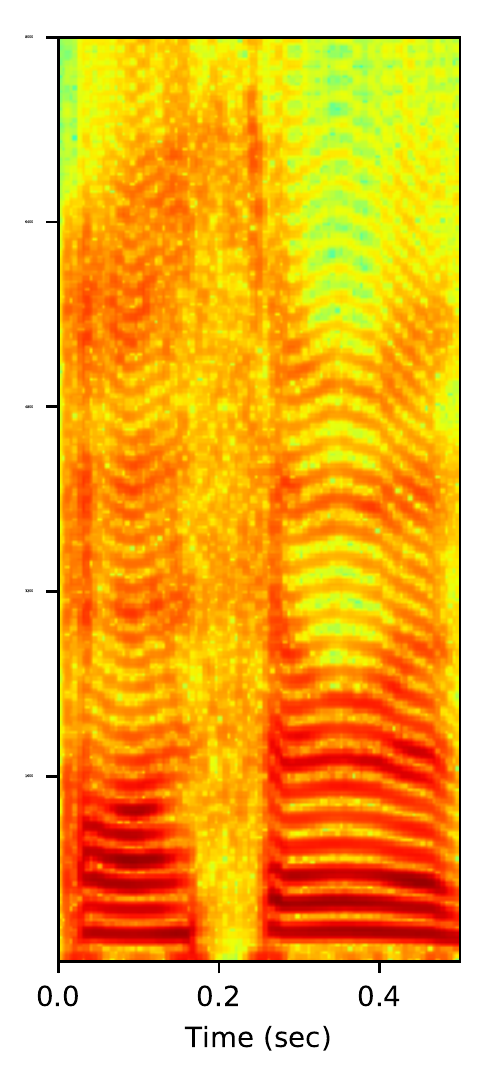}\label{fig:WORLD_stft_spectrogram}}
  \subfigure[STFT]{\includegraphics[height=5.2cm]{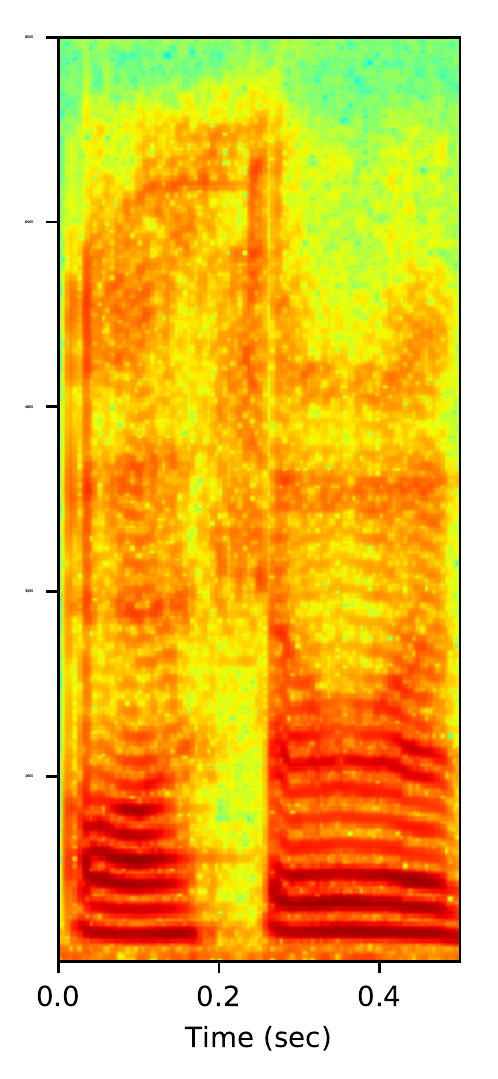}\label{fig:MSP-STFT_stft_spectrogram}}
  \subfigure[STFT+CWT]{\includegraphics[height=5.2cm]{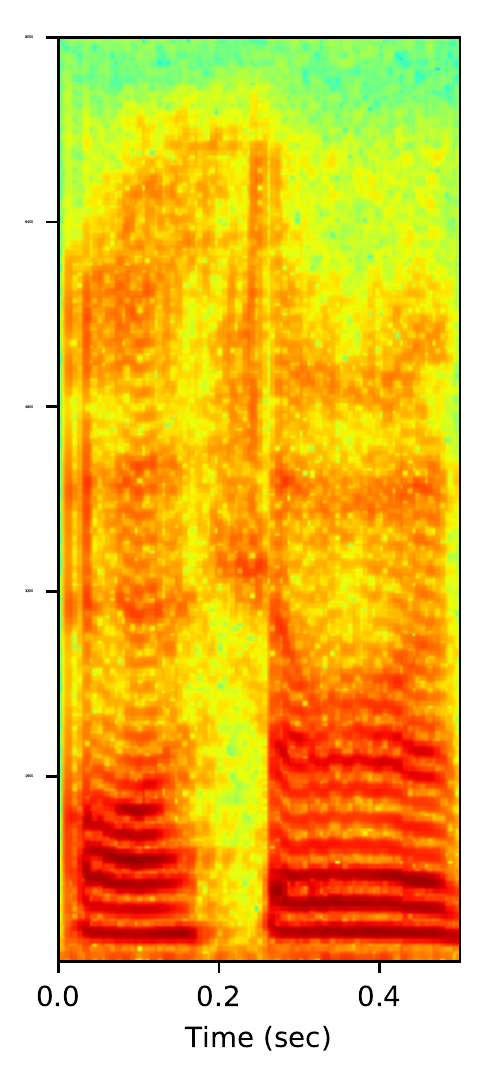}\label{fig:MSP-STFT+CWT_stft_spectrogram}}
  \subfigure[CWT]{\includegraphics[height=5.2cm]{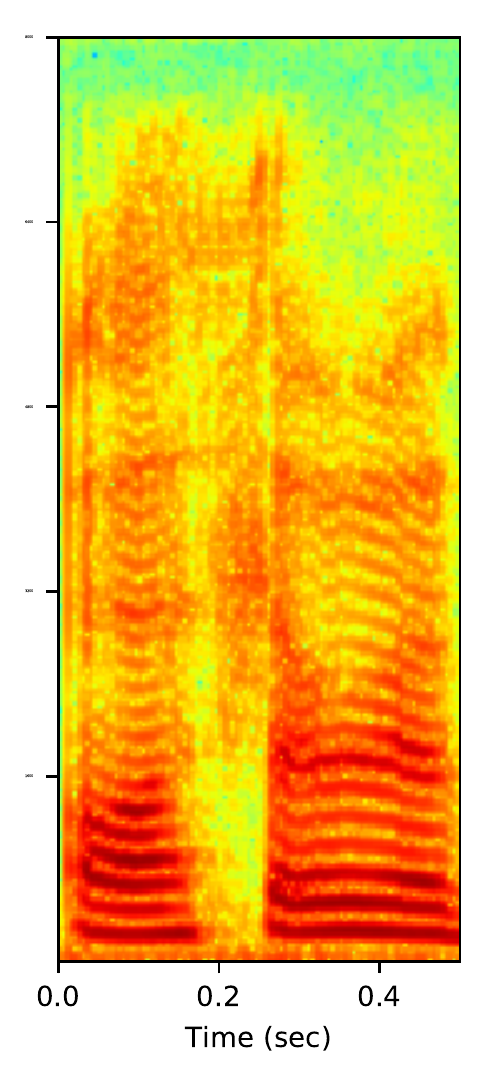}\label{fig:MSP-CWT_stft_spectrogram}}
  \caption{Parts of spectrograms obtained from natural speech and analysis-by-synthesis speech samples 
  }
  \label{fig:synth_spectrogram}
\end{figure*}

\subsection{Objective results}

In the experiments, analysis-by-synthesis speech samples, generated from the four vocoders listed in Table \ref{tab:method}, and natural speech samples (NAT) were evaluated. 

Parts of spectrograms obtained from natural speech and analysis-by-speech speech samples by utilizing the vocoders are shown in Fig.\ \ref{fig:synth_spectrogram}.  First, comparing spectrograms obtained from speech samples generated by utilizing neural vocoders with mel spectrograms as inputs [Figs. \ref{fig:MSP-STFT_stft_spectrogram}, \ref{fig:MSP-STFT+CWT_stft_spectrogram}, and \ref{fig:MSP-CWT_stft_spectrogram}], we can see that all spectrograms look similar. This means that CWT amplitude spectral loss can be used for training a high-performance speech waveform model, although the effectiveness of the combination of STFT and CWT spectral losses could not be observed from the spectrograms.

\subsection{Subjective results}
\begin{figure}[t]
  \centering
  \includegraphics[width=1.0\columnwidth]{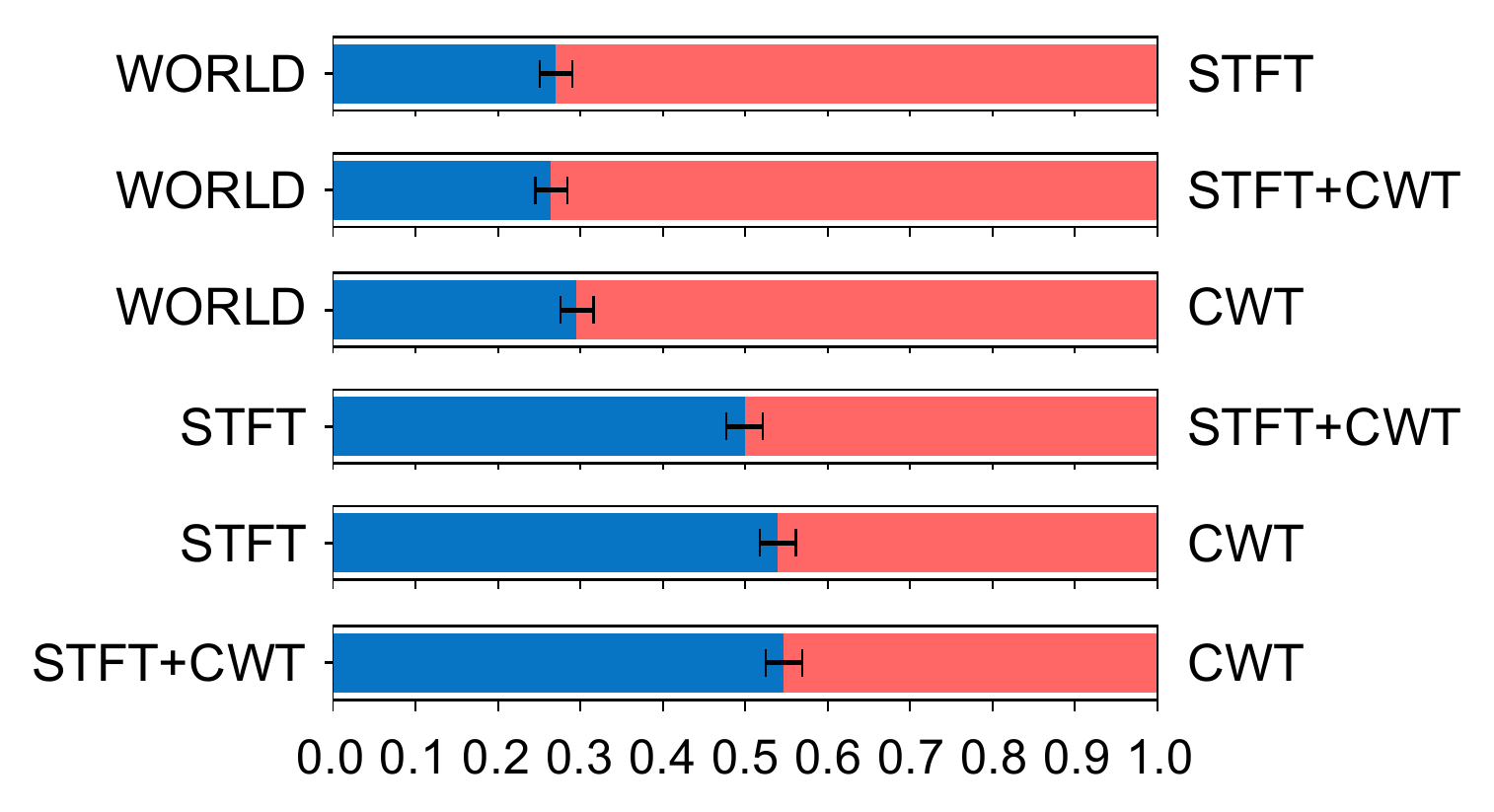}
  \caption{Subjective evaluation results. All pairs are statistically significant except for STFT vs.\ STFT+CWT.}
  \label{fig:SubjectiveTest_msp}
\end{figure}
%
%
%

The quality of the speech samples was evaluated in preference tests. This evaluation was carried out with a web-based interface on a crowdsourcing platform. On each web page, we presented two audio files and asked evaluators to state which sample sounds better. The evaluators were required to fully listen to the two samples before evaluating them. In total, 162 evaluators participated in the test. Each pair was evaluated $40$ times.

Six pairs were compared:   
\begin{itemize}
    \item World vocoder vs.\ Neural vocoder using the STFT loss
    \item World vocoder vs.\ Neural vocoder using the CWT loss
    \item World vocoder vs.\ Neural vocoder using the STFT and CWT loss
    \item Neural vocoder using the STFT loss vs.\ Neural vocoder using the STFT and CWT loss
    \item Neural vocoder using the STFT loss vs.\ Neural vocoder using the CWT loss
    \item Neural vocoder using the CWT loss vs.\ Neural vocoder using the STFT and CWT loss
\end{itemize}

Figure \ref{fig:SubjectiveTest_msp} shows the results of the preference tests. We can see that the neural waveform model using the proposed CWT loss was better than the standard deterministic World vocoder. We can also see that the proposed CWT loss was as good as our previous STFT-based loss from a comparison of ``CWT'' and ``STFT.'' However, unfortunately, we could not observe any benefits from combining the proposed CWT loss and STFT-based loss. One possible reason is that we used the AR-structure neural network for waveform modeling, and the impact of the conditional features may not have been significant. We should investigate the effectiveness of the proposed loss using a non-AR waveform model \cite{wang2018neural} as a next step.

\section{Conclusion}

In this paper, we extended STFT-based loss functions for training a neural speech waveform model and proposed a more generalized framework where we can use either STFT or CWT, or both of them. Experimental results showed that it is possible to train a high-quality model by using the proposed CWT spectral loss. Synthetic speech generated with the proposed loss sounds better than that generated with a standard deterministic speech vocoder and is as good as that with the STFT-based loss. 

However, in contrast to our expectation, combining the proposed CWT loss and STFT-based loss did not improve the quality of synthetic speech. Our future work includes investigating the effectiveness of the proposed loss for a non-AR waveform model \cite{wang2018neural}.

\section{Acknowledgements}
This work was partially supported by JST CREST Grant Number JPMJCR18A6 (VoicePersonae project), Japan and by MEXT KAKENHI Grant Numbers (16H06302, 16K16096, 17H04687, 18H04120, 18H04112, 18KT0051), Japan. The numerical calculations were carried out on the TSUBAME3.0 supercomputer at the Tokyo Institute of Technology.

\bibliographystyle{IEEEtran}
\bibliography{main}

\end{document}